# From Requirements to Test Cases: An NLP-Based Approach for High-Performance ECU Test Case Automation

Nikitha Medeshetty*, Ahmad Nauman Ghazi*, Sadi Alawadi[†], and Fahed Alkhabbas[‡] [§]
*Department of Software Engineering
Blekinge Institute of Technology, SE-37179 Karlskrona, Sweden
Email: nime22@student.bth.se, nauman.ghazi@bth.se
[†]Department of Computer Science
Blekinge Institute of Technology, SE-37179 Karlskrona, Sweden
Email: sadi.alawadi@bth.se
[‡]Department of Computer Science and Media Technology, Malmö University, Sweden
[§] Sustainable Digitalisation Research Centre, Malmö University, Sweden
Email: fahed.alkhabbas@mau.se

*Abstract*—Automating test case specification generation is vital for improving the efficiency and accuracy of software testing, particularly in complex systems like high-performance Electronic Control Units (ECUs). This study investigates the use of Natural Language Processing (NLP) techniques, including Rule-Based Information Extraction and Named Entity Recognition (NER), to transform natural language requirements into structured test case specifications. A dataset of 400 feature element documents from the Polarion tool was used to evaluate both approaches for extracting key elements such as signal names and values. The results reveal that the Rule-Based method outperforms the NER method, achieving 95% accuracy for more straightforward requirements with single signals, while the NER method, leveraging SVM and other machine learning algorithms, achieved 77.3% accuracy but struggled with complex scenarios. Statistical analysis confirmed that the Rule-Based approach significantly enhances efficiency and accuracy compared to manual methods. This research highlights the potential of NLP-driven automation in improving quality assurance, reducing manual effort, and expediting test case generation, with future work focused on refining NER and hybrid models to handle greater complexity.

## I. INTRODUCTION

High-performance Electronic Control Units (ECUs) are complex components that perform core functionalities in modern systems. Therefore, ECUs require rigorous testing to ensure their reliability, safety, and compliance with industry standards [15]. Traditional manual approaches of generating test cases for ECUs are time-intensive, prone to human errors, and require continuous updates to cope with evolving systems. Therefore, there is a need for approaches to automate the test cases generation process. High-performance electronic control units (ECUs) are crucial in modern vehicles, acting as the link between human input and machine response. This connection significantly impacts human-machine interaction (HMI). By automating the testing of ECU functions, we can enhance vehicle performance, optimize real-time decision-making, and improve the user experience. This ensures seamless communication between the driver and the vehicle's electronic systems. Natural Language Processing (NLP) is a promising technique that could be leveraged to achieve this goal by enabling the transformation of natural language requirements into structured and actionable test cases. Generally, NLP-driven approaches can enhance efficiency, accuracy, and scalability in test case specification [15]. Although NLP has been exploited for generating test cases automatically, existing approaches do not fully investigate the automated test case generation for ECUs [1].

To bridge this gap, in this paper, we propose an approach that exploits NLP, rule-based information extraction, and Named Entity Recognition (NER) techniques, which not only reduce human intervention and save time but also mitigate ambiguities inherent in natural language requirements. The main contribution of our paper is an approach that explores the application of these techniques to automate test case generation for high-performance ECUs. To validate the feasibility of our approach, we develop and run experiments use a dataset of 400 feature element documents from the Polarion tool to evaluate the approach's performance and suitability for varying levels of requirement complexity. The findings highlights the potential of NLP-driven automation in improving quality assurance, reducing manual effort, and expediting test case generation, with future work focused on refining NER and hybrid models to handle greater complexity.

## II. RELATED WORK

Product testing is used to identify errors and defects in developed systems. However, testing is a complex process and time-intensive, especially for big and complicated projects [1]. Thus, the written natural language requirement specifications can be transformed into semi-formal specifications utilizing a rule-based approach, enabling the creation of test cases through decision tables. Aoyama et al. [4] have observed that this approach helps to identify defects and significantly reduces the time required for human involvement in test design and execution. Given that the nature of test case development for software testing is time-consuming, various tools and methodologies have been introduced to automate this process. NLP further enhances this process by analyzing and extracting critical information from textual materials, such as user manuals or system documentation, to generate test cases covering diverse scenarios and edge cases can be created [8]. Moreover, leveraging NLP reduces human errors in the test cases generated manually and addresses ambiguities in requirements, making it an effective tool for automating test case generation [12].

Currently, practitioners in the automotive development sector mostly use a top-down approach, specifically behaviour-driven development (BDD). However, recent studies indicate that this approach faces limitations in deploying and evaluating automotive systems [13]. To address this limitation, a new approach has been proposed: integrating BDD with NLP techniques to enhance test scenario specifications. This approach combines top-down and bottom-up strategies to generate test cases from functional requirements, analyze textual specifications and integrate component-level behaviours into the test case specification process [1], [13]. However, Fischbach et al. [11] have explored automating the test case generation from requirements or acceptance criteria; achieving full automation remains challenging.

Extracting keywords from the text is imperative to generate test specifications. Various methods exist for this purpose, including part-of-speech tagging and morphological reduction for document preprocessing. These methods categorize words according to their grammatical roles (e.g., nouns, verbs) before implementing morphological transformations and utilizing algorithms such as Term Frequency-Inverse Document Frequency (TF-IDF), Latent Dirichlet Allocation (LDA), and Latent Semantic Indexing (LSI) for keyword extraction [6].

Another viable approach for keyword extraction is NER, combined with Rapid Automatic Keyword Extraction (RAKE), which improves keyword extraction efficiency, particularly for domain-specific data like High-Performance ECUs [10]. Custom NER models tailored to domain-specific entities, such as "signal" and "value," further enhance results. Research indicates that SVM-based named entity recognition systems outperform those based on Hidden Markov Models (HMMs). By leveraging a rich feature set and masking techniques, SVM-based systems yield reliable results across various named entity tasks [14]. Additionally, advancements in neural networks have improved text classification performance significantly, particularly models incorporating categorical metadata as supplementary information, utilizing product-related data. This integration has required modifications to several model components, including word embeddings and attention mechanisms, to enable result customization based on the metadata [10].

The aforementioned related works mainly focus on leveraging NLP to generate test cases and integrating various methods to enhance the quality of test specifications. However, there is a notable research gap regarding the impact of natural language processing on the generation of test case specifications, mainly when applied to the analysis requirements of High-Performance ECUs and the subsequent generation of test case specifications to improve testing quality.

## III. DATA ACQUISITION

The data used to validate this study was provided by Scania CV AB and includes manually written documentation for the High-Performance ECU, stored in the Polarion tool. The dataset comprises 400 feature element documents related to the High-Performance ECU. All requirements in the dataset are transformed into features, each with its detailed documentation, referred to as "feature element documents." These documents are structured as follows:

1) **Introduction:** provides an overview of the feature element, including its purpose, significance, and intended functionality. It serves as background information for the detailed content of the document.
2) **Parameters:** outlines the parameters essential for the feature element's operation, such as configuration settings, thresholds, and other critical variables influencing how the ECU manages engine performance.
3) **CAN Signal Inputs:** describes the Controller Area Network (CAN) signals the ECU requires to communicate with other sensors or ECUs within the truck. These signals provide real-time data for decision-making.
4) **CAN Signal Outputs:** details the output signals sent from the ECU to other components or systems. These outputs, derived from the ECU's processing, are critical in managing engine performance.
5) **Requirements:** specifies the prerequisites and criteria the feature element must meet, including functional requirements, performance metrics, safety standards, and regulatory compliance.

This structure ensures that each feature element document provides comprehensive and organized information for the High-Performance ECU. This organization plays a crucial role in creating accurate and detailed test case specifications for each feature element. Take into account that in this study, the term "**signal name**" refers to either the **CAN Signal name** or the **EOL Parameter name**, while the term "**signal value**" denotes the **CAN Signal value** or the **EOL Parameter value**.

## IV. OUR APPROACH

We exploit two technniques to generate test case specifications. The first one employs rule-based information extraction and utilizes manually defined rules and pattern-matching techniques to identify and extract relevant information, such as signal names and values, from feature element documents. Key operations for this technique include string manipulation for organizing and transforming textual data, pattern matching through regular expressions to detect defined patterns, and heuristic rules that capture specific structures within the text based on predefined guidelines. The second technique is called

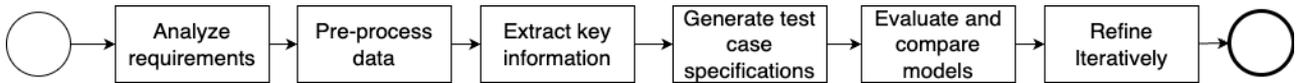

Fig. 1. The approach phases

Named Entity Recognition (NER), which uses machine learning models, including Support Vector Machine (SVM), Random Forest, Decision Tree, and Gradient Boosting classifiers, to identify specific entities within the text. The NER model is specifically designed to classify text into two predefined entities, signal and value. The use of machine learning enables this method to handle more complex and varied linguistic patterns compared to the rule-based approach. Named Entity Recognition (NER) is selected as the second method over other methods because to its ability to accurately extract and classify key entities from complex and unstructured technical documentation. High-performance ECUs involve extensive sets of structured and unstructured requirements documents. NER models are effective in generalizing across various types of ECU-related documents. Many of these documents contain ambiguous terms, NER understands context and can differentiate between similar but distinct entities.

Figure 1 shows the phases of our approach. In the *requirement analysis* phase, our approach focuses on identifying and extracting relevant section specifically, the requirements from feature element documents. Each feature element document comprises several sections, with the last section designated for the requirements. The requirements follow this format:

- If the signal ABC has a value of 0x1, then the telltale is active.
- If the signal ABC has a value of 0x0, then the telltale is inactive.

During the data preprocessing phase, the extracted feature elements are cleaned and organized using GUI automation, regular expressions (regex), and part-of-speech (POS) tagging. The output of this phase results in structured and cleaned data ready for analysis. After analyzing and pre-processing the requirements, they are converted into a tuple format. In this format, the first element represents the condition, and the second element denotes the action. The final output, after analysis and pre-processing, will be structured as tuples: *[(The signal ABC has value 0x1, telltale is active),( The signal ABC has value 0x0, telltale is Inactive)]*. The ECU can send various signals to different ECUs to facilitate communication. Each signal can have different values, which represent different messages, and these values are expressed in hexadecimal format. For example, signal ABC has a signal value of 0x1, which indicates the specific message that signal ABC is transmitting. The NER-based method will classify "ABC" as the signal name and "0X1" as the signal value. In contrast, the rule-based method has different rules based on the semantics of the requirements to extract both the signal name and signal value. In the key information extraction phase, our approach utilizes both rule-based methods and NER to extract critical details from the structured data, including signal names, values, and conditional statements. The possible values are 0x0, 0x1, and 0x2. Signal names and their corresponding values are extracted from a list of tuples using the two techniques: NER and rule-based methods. The extracted data is then stored as a list of dictionaries, where each dictionary consists of two key-value pairs: the first pair represents the signal name and its value, while the second pair outlines the expected output and the corresponding action of the condition. For instance, the list content could be as follows: {[{ABC:0x1, Expected output: telltale is active}, {ABC:0x0, Expected output: telltale is Inactive}, {ABC:0x2, Expected output: telltale is Inactive}]} Next, this list of dictionaries is transformed into a dataframe that contains test scenarios. Following this, a template is defined to rearrange the dataframe into test cases.

The final step involves generating test case specifications from the processed data. A predefined template structures each test case by focusing on individual signals and considering all possible signal-value combinations. Each test case includes a unique identifier, initial setup requirements, the specific signal to be tested, and the expected outcome. These test cases are then exported to an Excel file for further validation. This structured approach enables an organized and efficient generation of test cases, helping to ensure that all potential scenarios are adequately covered for reliable testing outcomes.

## V. EXPERIMENTAL SETTINGS

In this study, we perform extensive experiments using four machine learning (ML) models, i.e. SVM, Random Forest, Decision Tree, and Gradient Boosting Classifier, to evaluate the effectiveness of NER and Rule-Based Extraction for generating test case specifications. The dataset comprises 200 manually annotated entities, with detailed specifications of start and end positions for Signal and Value entities, as described in section III. To enhance robustness and minimize bias, a 10-fold cross-validation technique has been employed, ensuring statistical generalization and improving the learning process while maintaining consistent class distributions across training and testing sets. GUI automation techniques were utilized to address data extraction challenges and limitations using the Polarion tool. A custom Python script was developed to simulate user interactions, facilitating navigation through Polarion's interface and extracting relevant sections. Special attention was given to the IF-THEN conditions within the Requirements section, which were systematically stored as structured tuples. Signal names and corresponding values were identified and mapped using regular expressions and POS tagging, enabling efficient organization of the extracted information for subsequent analysis and processing by the ML models.

To evaluate the model's performance, we used different metrics, including accuracy, precision, recall, and the F1 score.

- **Accuracy** measures the ratio of correctly generated test cases to the total generated test cases:

$$Accuracy = \frac{TP + TN}{TP + TN + FN + FP} * 100\% \quad (1)$$

- **Precision** quantifies the proportion of correctly identified positive instances (true positives) among all instances identified as positive:

$$Precision = \frac{TP}{TP + FP} \quad (2)$$

- **Recall** (or Sensitivity) is the proportion of correctly identified positive instances (true positives) among all actual positive instances:

$$Recall = \frac{TP}{TP + FN} \quad (3)$$

- **F1 Score** is the harmonic mean of precision and recall, providing a balanced measure that accounts for both false positives and false negatives:

$$F1 = 2 \times \frac{Precision \times Recall}{Precision + Recall} \quad (4)$$

Where, True Positives (TP) are correctly identified positive cases, True Negatives (TN) are correctly identified negative cases, False Positives (FP) are negative cases wrongly classified as positive, and False Negatives (FN) are positive cases wrongly classified as negative.

## VI. RESULTS

The NER model was developed using a manually annotated dataset comprising 200 instances. Four algorithms were evaluated to determine their performance: SVM, Random Forest, Decision Tree Classifier, and Gradient Boosting Classifier. Among these, SVM demonstrated better performance with an accuracy of 77.3%, outperforming the other models, as presented in Table III. Comprehensive evaluation metrics, including precision, recall, and F1 score, were calculated for all algorithms. SVM achieved a precision of 0.816, recall of 0.773, and F1 score of 0.770, establishing its robustness for this NER task.

Here is an example of the requirements and the test case specification generated from these requirements. **Requirements**

- if T.DYN2.VehicleSpeedCheck has value 0X0 then MPG302 - Vehicle Speed Check is Not Active
- if T.DYN2.VehicleSpeedCheck has value 0X1 then MPG302 - Vehicle Speed Check is Active
- if T.DYN2.VehicleSpeedCheck has value 0X2 then MPG302 - Vehicle Speed Check is Not Active

This is the CAN-signal inputs table.

Figure 2 represents the test case specifications that are generated by analyzing the above requirements.

| CAN Signal | Range | Value table |
|---|---|---|
| VehicleSpeedCheck | 0-2 | 0X0, 0X1, 0X2 |

TABLE I
CAN-SIGNAL INPUTS

| Test case | Pre condition | T.DYN2.VehicleSpeedCheck | Expected Output |
|---|---|---|---|
| 1 | ECU is up and Running | Send CAN signal with value 0x0 | DDU shall set notification "MPG302 - Vehicle Speed Check" to "Not Active" |
| 1 | ECU is up and Running | Send CAN signal with value 0x1 | DDU shall request to put "MPG302 - Vehicle Speed Check" to "Active" |
| 1 | ECU is up and Running | Send CAN signal with value 0x0 | DDU shall set notification "MPG302 - Vehicle Speed Check" to "Not Active" |
| 2 | ECU is up and Running | Send CAN signal with value 0x0 | DDU shall set notification "MPG302 - Vehicle Speed Check" to "Not Active" |
| 2 | ECU is up and Running | Send CAN signal with value 0x1 | DDU shall request to put "MPG302 - Vehicle Speed Check" to "Active" |
| 2 | ECU is up and Running | Send CAN signal with value 0x0 | DDU shall set notification "MPG302 - Vehicle Speed Check" to "Not Active" |
| 2 | ECU is up and Running | Send CAN signal with value 0x1 | DDU shall request to put "MPG302 - Vehicle Speed Check" to "Active" |
| 2 | ECU is up and Running | Send CAN signal with value 0x2 | DDU shall set notification "MPG302 - Vehicle Speed Check" to "Not Active" |
| 2 | ECU is up and Running | Send CAN signal with value 0x1 | DDU shall request to put "MPG302 - Vehicle Speed Check" to "Active" |
| 2 | ECU is up and Running | Send CAN signal with value 0x0 | DDU shall set notification "MPG302 - Vehicle Speed Check" to "Not Active" |
| 2 | ECU is up and Running | Send CAN signal with value 0x1 | DDU shall request to put "MPG302 - Vehicle Speed Check" to "Active" |
| 2 | ECU is up and Running | Send CAN signal with value 0x3 | DDU shall set notification "MPG302 - Vehicle Speed Check" to "Not Active" |
| 2 | ECU is up and Running | Send CAN signal with value 0x1 | DDU shall request to put "MPG302 - Vehicle Speed Check" to "Active" |
| 2 | ECU is up and Running | Send CAN signal with value 0x0 | DDU shall set notification "MPG302 - Vehicle Speed Check" to "Not Active" |

Fig. 2. Test case specifications

To further validate SVM's performance, three Mann-Whitney U tests were conducted, each yielding a p-value below the 0.05 threshold. This confirms that SVM statistically outperforms the other algorithms at a 95% confidence level.

### A. Testing Rule-Based Method Against Manual Method

**Null Hypothesis** ($H_0$): The Rule-Based Method's accuracy is less than or equal to that of the Manual Method.

**Alternative Hypothesis** ($H_1$): The Rule-Based Method's accuracy exceeds that of the Manual Method.

A Z-Test was performed to evaluate this hypothesis. The results showed a p-value of 0.8060, significantly exceeding the 0.05 significance level. Thus, the null hypothesis could not be rejected, indicating insufficient evidence to support the superiority of the Rule-Based Method.

### B. Experimental Workflow

The experiment was executed in three primary stages:

**1. Data Extraction and Preprocessing:** Data was sourced from feature element documents and preprocessed to ensure consistency and relevance.

**2. Information Extraction:** Essential information, such as signal names and values, was extracted using two approaches: NER and Rule-Based Information Extraction.

**3. Test Case Specification Generation:** Test case specifications were generated based on the extracted information. While Steps 1 and 3 remained consistent, Step 2 utilized distinct approaches for NER and Rule-Based methods.

Feature elements were categorized based on the complexity of their signal composition:

- **Category 1:** Single signal feature elements.
- **Category 2:** Feature elements with up to four signals.
- **Category 3:** Feature elements with more than four signals.

### C. Model Performance Across Categories

The accuracy of the Rule-Based method demonstrated a notable dependency on signal complexity. For Category 1, which included single-signal feature elements, the Rule-Based model achieved an accuracy of 95%. In Category 2, involving up to four signals, accuracy declined to 75%. Category 3, characterized by feature elements with more than four signals, exhibited the lowest accuracy at 60%. These findings are summarized in Table II.

| Category | Description | Accuracy |
|---|---|---|
| Category 1 | Feature elements with a single signal | 95% |
| Category 2 | Feature elements with up to four signals | 75% |
| Category 3 | Feature elements with more than four signals | 60% |

TABLE II
ACCURACY ACROSS FEATURE ELEMENT CATEGORIES

The results indicate an inverse relationship between signal complexity and model accuracy in the Rule-Based method. While high accuracy was observed for simple single-signal feature elements, the performance declined with increased signal complexity in multi-signal feature elements. This highlights the need for enhanced algorithms to address the challenges posed by complex scenarios.

*D. Feedback Analysis and Model Refinement*

Feedback from a team evaluation validated the model's reliability for single-signal feature elements, underscoring its effectiveness in generating accurate test case specifications for expected ECU behavior. However, the team emphasized the need for refinement in handling multi-signal scenarios. For feature elements incorporating more than one signal, the current model logic requires further development to enhance accuracy and completeness.

Overall, the model exhibits strong performance for simpler cases, achieving baseline accuracies of 95% for single-signal feature elements and 75% for elements with up to four signals. As summarized in Table III, SVM emerged as the most effective algorithm, establishing a robust foundation for NER-based approaches in straightforward scenarios. Figure 3 illustrates the confusion matrices for all evaluated algorithms. Future work will focus on improving the model's capability to manage the complexity of multi-signal feature elements, ensuring comprehensive test case generation for diverse scenarios.

| Algorithm | Accuracy | Precision | Recall | F1 Score |
|---|---|---|---|---|
| SVM | 0.773 | 0.816 | 0.773 | 0.770 |
| Random Forest | 0.408 | 0.263 | 0.408 | 0.316 |
| Decision Tree Classifier | 0.182 | 0.165 | 0.174 | 0.163 |
| Gradient Boosting Classifier | 0.170 | 0.155 | 0.170 | 0.160 |

TABLE III
PERFORMANCE METRICS OF EVALUATED ALGORITHMS

## VII. DISCUSSION

This study explored the comparative efficacy of Rule-Based and Named Entity Recognition (NER) techniques in automating test case specification generation. The findings indicate that the Rule-Based Method excels in simpler scenarios, delivering exceptional accuracy of 95% for single-signal feature elements and 75% for elements with up to four signals. These results underscore the deterministic strengths of the Rule-Based Method, particularly in cases where predefined rules can reliably model the expected behavior. However, as feature complexity increases, the Rule-Based Method's performance diminishes significantly, with accuracy dropping below 60% for multi-signal feature elements. This decline underscores the inherent limitations of deterministic approaches when faced with high-dimensional and intricate data inputs.

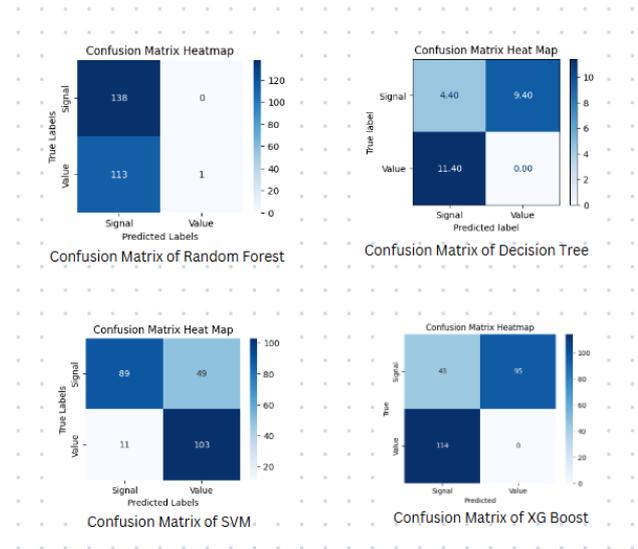

Fig. 3. Confusion matrix of evaluated algorithms

In contrast, the NER approach, leveraging machine learning algorithms such as SVM, Random Forest, Decision Tree, and Gradient Boosting, showcased variable performance across tasks. Among these, SVM emerged as the most robust performer, achieving an accuracy of 77.3% following meticulous feature extraction and hyperparameter optimization. This highlights its potential for managing high-dimensional data in NER contexts. Nevertheless, the overall performance of NER techniques was less effective than the Rule-Based Method in simpler scenarios. Challenges associated with natural language processing (NLP) tasks, including domain-specific complexities and the inherent variability of language, contributed to this discrepancy. Notably, the poor performance of Gradient Boosting, with an accuracy of 17%, further exemplifies the inconsistency of NLP-driven approaches in domain-specific applications.

Both methods face challenges when addressing complex, multi-signal feature elements. While the Rule-Based Method struggles with capturing the intricate interactions among multiple signals, NER approaches are hindered by increased task complexity in entity recognition and information extraction. These findings suggest the need for innovative solutions, such as hybrid models that integrate the deterministic reliability of rule-based systems with the adaptability of machine learning algorithms.

Statistical analyses support these observations. Hypothesis testing confirmed the Rule-Based Method's significant improvement over manual methods, with a p-value of 0.037. Conversely, the NER approach failed to demonstrate statistically significant improvements over manual methods, indicated by a p-value of 0.806. This reinforces the necessity for further optimization of NER models to establish their viability in automated test case generation.

To advance the utility of NER methods, this study recommends increasing the diversity and volume of training data, particularly for complex feature elements, to enhance model generalization and reliability. Additionally, developing hybrid approaches that combine rule-based logic with machine learning techniques may offer a balanced solution, harnessing the strengths of both methodologies. From a practical standpoint, the Rule-Based Method demonstrates a transformative potential in industrial settings by reducing test case generation time from approximately 24 hours to just 5 minutes per case. Such efficiency gains promise significant improvements in productivity and responsiveness, enabling software testing teams to adapt swiftly to evolving requirements and deliver high-quality outputs.

## VIII. CONCLUSION

This study presents a novel approach for automating the creation of test case specifications for high-performance ECUs, leveraging two natural language processing techniques: the Rule-Based Method and Named Entity Recognition (NER). The primary objective is to streamline test case generation by extracting critical information from feature element documents, thereby enhancing both accuracy and efficiency compared to traditional manual methods. The evaluation, conducted on a dataset comprising approximately 200 sentences from 60 feature element documents, provided insights into the effectiveness of these techniques.

The Rule-Based Method, based on predefined linguistic rules, emerged as the most efficient and accurate approach for generating precise test case specifications. Its performance was particularly noteworthy in handling less complex requirements, where it successfully transferred information from feature documents to test cases with an impressive accuracy of up to 95%. Conversely, the NER approach, implemented with machine learning algorithms such as SVM, Random Forest, Decision Tree Classifier, and Gradient Boosting Classifier, demonstrated variable performance. Among these, SVM achieved the highest accuracy at 77.3%, reflecting its potential in managing high-dimensional data. However, NER, in its current implementation, fell short of matching the accuracy and practicality of the Rule-Based Method.

The findings of this study show that the Rule-Based Method's practicality and reliability, especially for straightforward requirements, making it a superior alternative to both NER and traditional manual methods for test case generation. Nonetheless, the study also highlights the need for further optimization of the NER approach to better address complex feature elements. Enhancing the adaptability and precision of NER models could pave the way for broader applicability in automating software testing tasks.

In conclusion, this research demonstrates the significant potential of natural language processing techniques in transforming software testing processes. By automating test case generation, the proposed approaches can substantially improve accuracy and reduce the time and effort required for manual testing. Future work will focus on refining NER methodologies and exploring hybrid models that integrate the deterministic strengths of rule-based systems with the adaptability of machine learning algorithms, ensuring comprehensive and robust solutions for diverse software testing scenarios.